\documentclass{article}
\usepackage{arxiv}

\author{
        Andrew Mellor* and Angelica Grusovin \\
        Mathematical Institute\\
		University of Oxford\\
		*\texttt{mellor@maths.ox.ac.uk} 
}
\date{\today}

\usepackage{hyperref}
\usepackage{graphicx}
\usepackage[format=plain,justification=centerlast]{caption}
\usepackage{subcaption}
\usepackage{xcolor}
\usepackage{dsfont}
\usepackage{booktabs}
\usepackage{bm}
\usepackage{amsmath, amsfonts}
\usepackage[numbers]{natbib}

\DeclareMathOperator{\tr}{tr}

\begin{document}

\title{Graph Comparison via the Non-backtracking Spectrum}

\maketitle

\begin{abstract}
The comparison of graphs is a vitally important, yet difficult task which arises across a number of diverse research areas including biological and social networks.
There have been a number of approaches to define graph distance however often these are not metrics (rendering standard data-mining techniques infeasible), or are computationally infeasible for large graphs.
In this work we define a new metric based on the spectrum of the non-backtracking graph operator and show that it can not only be used to compare graphs generated through different mechanisms, but can reliably compare graphs of varying size.
We observe that the family of Watts-Strogatz graphs lie on a manifold in the non-backtracking spectral embedding and show how this metric can be used in a standard classification problem of empirical graphs.
\end{abstract}

\section{Introduction}

Comparing graph structures is a fundamental task in graph theory.
In particular the need to identify similar graph structure arises in order to determine common function, behaviour, or generative process.
This need is universal across disciplines and applications range from image processing \cite{conte2004thirty}, chemistry \cite{allen2002cambridge, kvasnivcka1991reaction}, and social network analysis \cite{koutra2013deltacon, lyzinski2016graph}.

For small graphs, identifying two identical graph structures (up to an isomorphism, or relabelling of vertices) is a trivial and computationally feasible task due to the limited number of possible configurations.
For graphs with four vertices there are six possible graph configurations.
However, for graphs with ten vertices there are $11,716,571$ possible configurations.
Although there is not a closed formula for the number of graph configurations with $n$ vertices, it is clear that it soon becomes intractable to enumerate or compare large graphs through brute-force.

The problem of determining whether two graphs are isomorphic is equivalent to finding a permutation matrix $P$ such that $AP = PB$ where $A$ and $B$ are the adjacency matrices of the two graphs.
Despite the difficulty in the task\footnote{
    It is currently unknown as to whether this problem lies in the class of P or NP-hard problems \cite{babai2016graph}
}
there have been a number of attempts to define graph distance by minimising $||AP - PB||$ for a suitably defined matrix norm $||\cdot||$.
The matrices $A$ and $B$ can also take different forms to shift from a local perspective (adjacency matrix) to consider global properties, where $A$ and $B$ capture the number of paths between vertices.
Examples of such distances include the chemical distance \cite{kvasnivcka1991reaction}, CKS distance \cite{chartrand1998graph}, and edit distance \cite{garey2002computers, sanfeliu1983distance}.
Recent work has focused on reducing the space over which to find a permutation matrix \cite{bento2018family} which makes the problem tractable, although still computationally restrictive for large graphs.

Beyond scalability issues, one of the drawbacks of these methods is that they require graphs to be of the same order (i.e. the same number of vertices).
For many applications this may be desirable, especially where the addition of a single vertex or edge can have drastic consequences on graph function (in graphs of chemical compounds for example).
However for other applications we may be interested in the structure at a coarser level.
For example, a ring graph of $100$ vertices is structurally similar to a ring graph of $200$ vertices, however the two cannot be compared using traditional isomorphism arguments. 

Contrasting approaches use a number of graph properties, or \emph{features}, to characterise graph structure \cite{wegner2017identifying, ferrer2010generalized, klau2009new, riesen2007graph, onnela2012taxonomies}.
The most promising of these make use of the graph spectrum, either of the adjacency matrix directly, or of a version of the graph Laplacian \cite{de2016spectral, wilson2008study,zhu2005study,elghawalby2008measuring, luo2003spectral}.
The spectra of graphs or graph operators are useful since the eigenvalues and eigenvectors characterise the topological structure of a graph in a way which can be interpreted physically.
Eigenvalues of the graph Laplacian describe how a quantity (information, heat, people, etc.) localised at a vertex can spread across the graph.
These eigenvalues also dictate the stability of dynamics acting across the graph \cite{newman2010networks}.
Beyond the Laplacian, recent work has investigated the spectral properties of higher-order operators such as the non-backtracking graph operator (described in the next section) \cite{torres2018graph}.
For a more detailed (but not complete) examination of graph distances we refer the reader to a recent survey \cite{donnat2018tracking}.

In this article we present a new method to compare graph structure using the distribution of eigenvalues of the non-backtracking operator in the complex plane.
In Section~\ref{sec:nonbacktracking} we describe the non-backtracking operator and investigate some of its spectral properties before introducing the distributional non-backtracking spectral distance (d-NBD) in Section~\ref{sec:distance}.
We show empirical results for both synthetic and real graphs in Section~\ref{sec:results} before discussing the significance of the results and future research in Section~\ref{sec:conclusions}.

\section{The Non-backtracking Operator and Spectrum}
\label{sec:nonbacktracking}

Consider a simple undirected binary graph $G=(V,E)$ where $V \subset \mathbb{N}$ is a set of vertices and $E \subset V^2$ is a set of edges.
Here we use the word \emph{simple} to mean that the graph contains no self-loops or multiple edges between vertices.
Let $n=|V|$ be the number of vertices and $m=|E|$ be the number of edges.

Spectral analysis of graphs is typically conducted using the Laplacian matrix $L=A-D$ where $A$ is the adjacency matrix which encodes $G$, and $D$ is a diagonal matrix of vertex degrees with $D_{ii} = \sum_j A_{ij}$ and zeros elsewhere.
In this work we chose instead to use a different linear graph operator, namely the non-backtracking operator.
The non-backtracking (or Hashimoto) matrix $B$ is a $2m \times 2m$ matrix defined on the set of directed edges of $G$.
Here each undirected edge $u \leftrightarrow v$ is replaced by two directed edges $u \to v$ and $v \to u$.
The non-backtracking matrix is then given by
\begin{align*}
    B_{(u\to v),(x \to y)} = \begin{cases}
        1 & \mbox{ if } x=v \mbox{ and } y \neq u \\
        0 & \mbox{ otherwise.}
    \end{cases}    
\end{align*}
The non-backtracking matrix is asymmetric and the entries of $B^k$ describe the number of non-backtracking walks of length $k$ across $G$.
Of special note is $(B^k)_{ii}$ which counts the non-backtracking cycles of length $k$ which start and end at edge $i$\footnote{Non-backtracking does not mean that a walk only visits distinct vertices, only that it does not immediately return to a vertex it has just arrived from.}.
The total number of non-backtracking cycles of length $k$ is therefore captured in the trace of $B^k$.

\subsection{Spectral Properties}

\begin{figure}
	\centering
	\includegraphics[width=\linewidth]{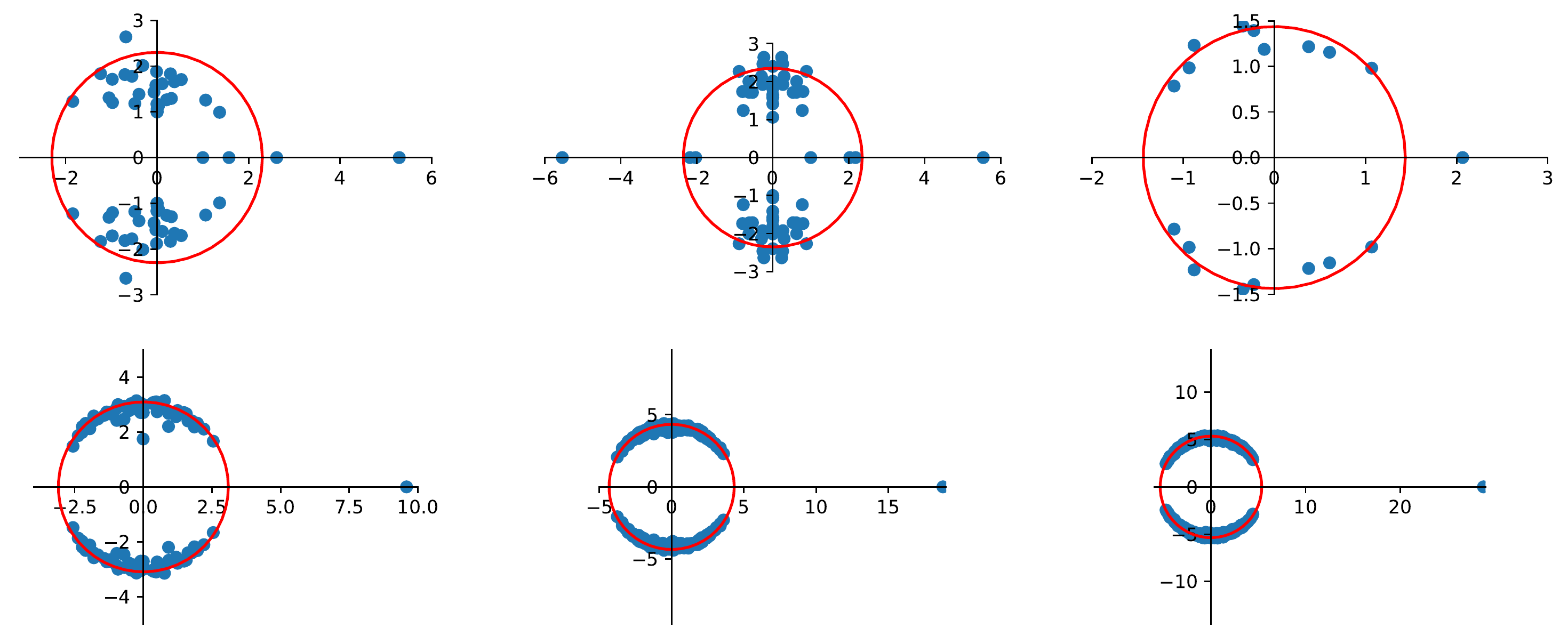}
	\caption{
        The non-backtracking spectrum of real and synthetic networks in the complex plane.
        Each eigenvalue $\lambda_k = \alpha_k + \beta_k i$ is represented by the point $(\alpha_k, \beta_k)$. 
		(Top, left to right) The Karate Club graph \cite{zachary1977information}, Davis Southern Women graph \cite{davis2009deep},  and Florentine Families graph \cite{breiger1986cumulated}.
        (Bottom, left to right) An Erd\H{o}s-R\'enyi graph with edge connection probability $p=0.2$ for $50,100$ and $150$ vertices.
        The red circle marks the curve $\alpha^2 + \beta^2 = \sqrt{\rho(B)}$ where $\rho(B)$ is the spectral radius of $B$. 
	}
	\label{fig:spectrum_examples}
\end{figure}

The spectral properties of the non-backtracking matrix are well understood, especially for regular graphs \cite{mckay1981expected, angel2015non, saade2014spectral}.
As $B$ is asymmetric the eigenvalues can also take complex values.
In Figure~\ref{fig:spectrum_examples} we show the spectrum in the complex plane of three real-world graphs and three synthetic graphs of varying size.

The first observation is that the bulk of the eigenvalues $\lambda_k$ lie within the disk $|\lambda_k| < \sqrt{\rho(B)}$ where $\sqrt{\rho(B)} = \max_k|\lambda_k|$ is the spectral radius of $B$.
This is commonly used as a heuristic for the bulk of the spectrum, however in particular cases more is known.
For random regular graphs of degree $d$ there are $n$ pairs of conjugate eigenvalues which lie exactly on a circle of radius $\sqrt{d-1}$, and for a stochastic block model the expected magnitude of the eigenvalues can be shown to be less than or equal to $\sqrt{\rho(B)}$ in the limit $n \to \infty$ \cite{krzakala2013spectral}.

Another property of the spectrum of $B$ is that any tree structure (whether connected to the graph, or disconnected) contributes zero eigenvalues to the spectrum.
This is a consequence of any non-backtracking walk becoming `stuck' at the leaves of the tree.
Furthermore any unicyclic component gives rise to eigenvalues in $\{0\} \cup \{\lambda: |\lambda|=1\}$.
These two properties are easily explained as we make the connection between the number of non-backtracking cycles in the graph and the eigenvalues of $B$.

For any matrix $A$ the trace of $A^k$ can be calculated as 
\begin{align*}
    \tr{A^k} = \sum_{i=1}^{n} \lambda_i^k
\end{align*}
where $(\lambda_i)_{i=1}^{n}$ are the eigenvalues of $A$.
This result is trivial for symmetric matrices which can be diagonalised, however the result also holds for asymmetric matrices.
We can therefore write
\begin{align}
    \tr{B^k} = \sum_{i=1}^{2m} \lambda_i^k,
    \label{eqn:trace_eigs}
\end{align}
noting that $\tr{B^k}$ captures a count of all non-backtracking cycles of length $k$.
This means that the eigenvalues of $B$ encode all information regarding the number of non-backtracking cycles in $A$.
Using this relation it is then clear that tree structures contribute zero eigenvalues since they cannot be part of a non-backtracking cycle.

The last property of the spectrum we detail is derived from the Ihara determinant formula \cite{bass1992ihara, hashimoto1989zeta, horton2006zeta}.
For any finite and undirected graph
\begin{align*}
    \det(B - \lambda I_n) = (\lambda^2-1)^{|E|-|V|}\det(Q_\lambda) 
\end{align*}
where $Q_\lambda = (D-I_n) - \lambda A + \lambda^2 I_n$ and $D$ is the matrix with vertex degrees on the diagonal with zeros elsewhere.
Using a linearisation of the quadratic polynomial $Q_\lambda$\footnote{This is in the form of a quadratic eigenvalue problem (QEP).} one can show that the $2n \times 2n$ matrix
\begin{align}
B' = \left(\begin{array}{cc}A & I-D \\ I & 0 \end{array}\right)
\end{align}
shares the same eigenvalues as $Q_\lambda$, i.e.,
\begin{align*}
    \det(Q_\lambda) = \det(B'- \lambda I_n).
\end{align*}
Therefore all eigenvalues of $B$ save for $|E|-|V|$ eigenvalues of $\pm 1$ are eigenvalues of the smaller matrix $B'$.
This result has very practical implications for calculating the spectrum of $B$ given that $n$ can often be substantially smaller than $m$.
The matrix $B'$ provides a means to calculate a significant proportion of the spectrum of $B$ in a fashion that scales linearly with the total number of vertices.

\section{Distributional Non-backtracking Spectral Distance}
\label{sec:distance}

Much like the spectrum of the graph Laplacian has found use in community detection \cite{donetti2004detecting}, dynamics on graphs \cite{newman2010networks} and graph clustering \cite{ren2011graph} our aim is to exploit the non-backtracking spectrum to differentiate between graph structures.
Recent work \cite{constantine2018marked, torres2018graph} has shown that the two-cores extracted from two graphs are isomorphic when the set of all non-backtracking cycles in each graph are equal (this is referred to as the length spectrum).
This means that should we be able to enumerate all possible non-backtracking cycles we can effectively compare two graph structures.
In practice enumerating all possible cycles of all possible lengths is infeasible, and even so it remains unclear how two such sets should be compared.

\citet{torres2018graph} address this issue by considering a `relaxed' length spectrum, calculating the \emph{number} of cycles of length $k$ as opposed to the set of cycles themselves.
As shown in \eqref{eqn:trace_eigs} the number of these cycles is captured completely by the spectrum of $B$.
Based on this argument the distance between the number of cycles (and therefore eigenvalues) should provide a reasonable approximation to the graph isomorphism.

In \cite{ren2011graph} the number of non-backtracking cycles of length $k$, $c_k$ are used directly as an embedding of the graph.
Each graph is embedded as a vector $\vec{v}_1 = [c_3, c_4, c_5, c_6, c_7, \ln(c_{2m})]$ with the graph distance subsequently defined as the Euclidean distance between vectors.
These distances correlate well with the graph edit distance and perform marginally better than a truncated Laplacian spectrum embedding when applied to a computer vision task.

In contrast, the approach in \cite{torres2018graph} uses the largest $k$ eigenvalues of $B$ (in order of magnitude) to construct a feature vector.
In particular for an ordered sequence of eigenvalues $(\lambda_i)_{i=1}^{2m}$ the graph is embedded as 
\begin{align}
    \vec{v}=[\alpha_1, \dots, \alpha_k, \beta_1, \dots, \beta_k],
    \label{eqn:nbd_embedding}
\end{align}
where $\lambda_s = \alpha_s + \beta_s i$ and again defining graph distance using the Euclidean distance between embeddings.
Taking the largest eigenvalues is an intuitive step as these contribute the most to the cycle counts.
There are however many instances where the magnitude of the eigenvalues are approximately equal (seen in Figure~\ref{fig:spectrum_examples}) and so the method is heavily sensitive to the choice of $k$ and the method of eigenvalue calculation.
This is especially problematic for a $d$-regular random graph where all non-real eigenvalues of $B$ are conjugate pairs on the circle of radius $\sqrt{d-1}$.

The two previous approaches to non-backtracking distance are limited in their ability compare graphs of different size.
Consider two graphs of sizes $n_1$ and $n_2$.
Naturally as we increase the number of vertices and edges the number of possible cycles will increase and therefore any cycle counting distance will diverge as ${|n_1 - n_2| \to \infty}$.
Furthermore, in order to compare the eigenvalues of the non-backtracking matrices directly we are restricted to choosing the largest $\min\{2n_1, 2n_2\}$ eigenvalues.
We are therefore neglecting a potentially significant part of the spectrum for the larger graph, i.e. we are using a truncated spectrum.
The issues surrounding the truncated spectrum become clear for $d$-regular random graphs.
Since the bulk of all eigenvalues like on the circle $|\lambda|=\sqrt{d-1}$ we cannot order them by magnitude.
We therefore need a secondary ordering, say in descending order of $\Re(\lambda)$.
The full spectrum will cover all eigenvalues on the circle, however the truncated spectrum will lie on an arc of $|\lambda|=\sqrt{d-1}$.
An effective comparison is therefore not possible.

This motivates us to consider a property of the spectrum that can be compared across graphs of different size, namely the spectral distribution.
In this sense we hope to cluster graphs which share the same properties of generating mechanism, regardless of size.
For computational efficiency we consider the eigenvalues of
\begin{align*}
B' = \left(\begin{array}{cc}A & I-D \\ I & 0\end{array}\right)
\end{align*}
where now $A$ is the two-core of the original graph (vertices of degree one are iteratively removed), and $D$ is the corresponding degree matrix.
This means that we are operating on a reduced spectrum with no zero eigenvalues and $|E|-|V|$ eigenvalues of $1$ or $-1$ omitted.
The spectral radius of $B'$ scales with the largest vertex degree, however the corresponding eigenvalues are not localised around high degree vertices to the same extent as they are with the graph Laplacian.
This is due to walks on $B'$ not being permitted to return to high degree vertices immediately after leaving them.
This suggests a rescaling of the eigenvalues
\begin{align*}
    |\lambda_i| \to \log_{\rho(B)}|\lambda_i|.
\end{align*}
The rescaled eigenvalues $(\hat{\lambda})_{i=1}^{2n}$ then lie exclusively in the disk $|\hat{\lambda}| \le 1$ with the bulk of the eigenvalues distributed in $|\hat{\lambda}| \le \frac{1}{2}$.

\begin{figure}
    \centering
    \includegraphics[width=\linewidth]{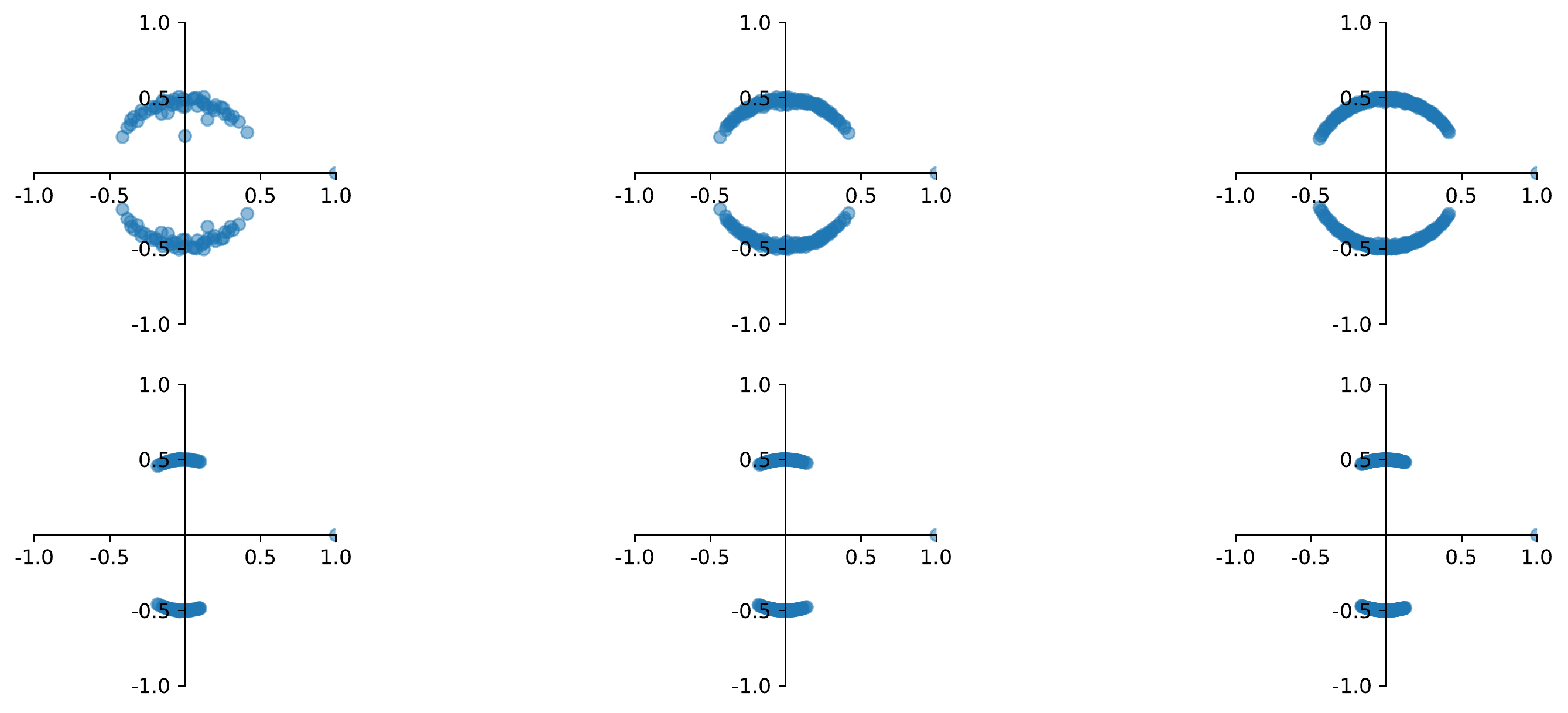}
    \caption{
        The rescaled non-backtracking spectrum for Erd\H{o}s-R\'enyi graphs of size $50,100,150$ (left to right) and edge connection probability $0.2$ (top) and $0.8$ (bottom).
        The rescaled spectrum is persistent as the number of vertices in the graph increases, however for small graphs there are finite size effects.
        In this example the two parameter regimes are distinguished by the range over which the argument of the eigenvalues take.
    }
    \label{fig:spectrum_examples_rescaled}
\end{figure}

Figure~\ref{fig:spectrum_examples_rescaled} shows the rescaled spectrum for Erd\H{o}s-R\'enyi graphs of varying size with edge connection probabilities $p=0.2$ (top) and $p=0.8$ (bottom).
Here we can clearly distinguish between the two parameter regimes and we see that these distributions are consistent across graphs irrespective of size.

We capture the distribution of the eigenvalues through the empirical cumulative spectral density given by
\begin{align}
    F(r, \theta) = \frac{1}{2n} \sum_{i=1}^{2n} \mathds{1}_{\{|\hat{\lambda}_i| \leq r\}} \mathds{1}_{\{0 \leq \arg(\hat{\lambda}_i) \leq \theta \}}
    \label{eqn:empirical_distribution}
\end{align}
This is defined on $r \in [0,1]$, $\theta \in [0, \pi]$ and for $n>0$.
A special case to mention is when the two-core of a graph is empty.
In this case we set $F(r,\theta)=0$ $\forall r,\theta$ by convention\footnote{
    Alternatively we could set $F(r,\theta) = \infty$ so that all distances to the empty graph are infinite.
}.
Since the complex eigenvalues of $B$ (and therefore $B'$) are conjugate pairs we gain no further information by considering the lower half of the complex plane and is hence omitted.
This naturally leads us to define the distributional non-backtracking spectral distance (d-NBD) as the distance between empirical spectral densities.
For two graphs $G_1$ and $G_2$ with rescaled spectral densities $F_1$ and $F_2$ respectively, the d-NBD is given by
\begin{align}
    d(G_1,G_2) = \frac{1}{\pi} \left(\int_0^{\pi} \int_0^1 |F_1(r,\theta) - F_2(r,\theta)|^2 dr d\theta \right)^{\frac{1}{2}},
    \label{eqn:dnbd}
\end{align}
i.e. $d(G_1,G_2)$ is proportional to the $p$-norm $||F_1 - F_2||_p$ with $p=2$.

The d-NBD satisfies many desirable properties.
It is non-negative, symmetric, and satisfies the triangle inequality (via the Minkowski inequality).
While it is evident that $d(G,G)=0$ it is not the case that $d(G_1, G_2)=0$ implies that $G_1=G_2$, only that they share the same two-core.
The d-NBD is therefore a \emph{pseudometric}.

Finally we define a $k^2$-dimensional embedding of a graph through a discretization of the empirical spectral density.
Here a graph is represented by 
\begin{align}
\label{eqn:embedding}
  \vec{v} = [&{F}(r_1,\theta_1), \dots, {F}(r_k, \theta_1),\\ \nonumber
   &{F}(r_1,\theta_2), \dots, {F}(r_k, \theta_2),\\ \nonumber
   &\phantom{{F}(r_1,\theta_2),} \dots \\ \nonumber
   &{F}(r_k, \theta_1) \dots, {F}(r_k, \theta_k)]
\end{align}
where $(r_1,\theta_1) = (0,0)$ and $(r_k,\theta_k) = (1,\pi)$.
This embedding is useful in the next section to visualise the d-NBD by plotting the graph embeddings in low-dimensional space.

\section{Results}
\label{sec:results}

In this section we give a number of results from both synthetic and real-world graph examples.

\subsection{Synthetic Graphs}

\begin{figure}
    \centering
    \begin{subfigure}{0.49\linewidth}
        \centering
        \includegraphics[width=\linewidth]{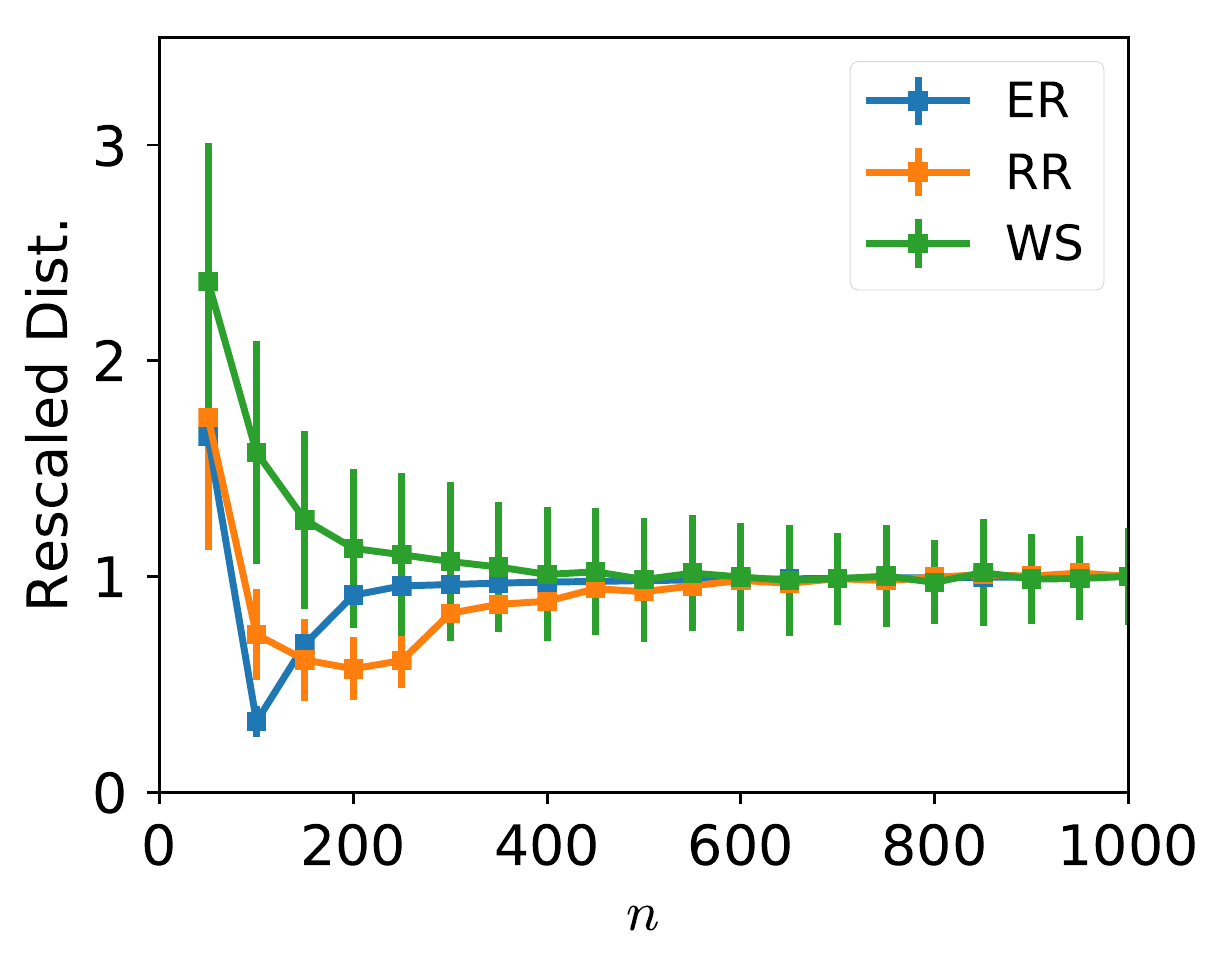}
        \caption{}
    \end{subfigure}
    \begin{subfigure}{0.49\linewidth}
        \centering
        \includegraphics[width=\linewidth]{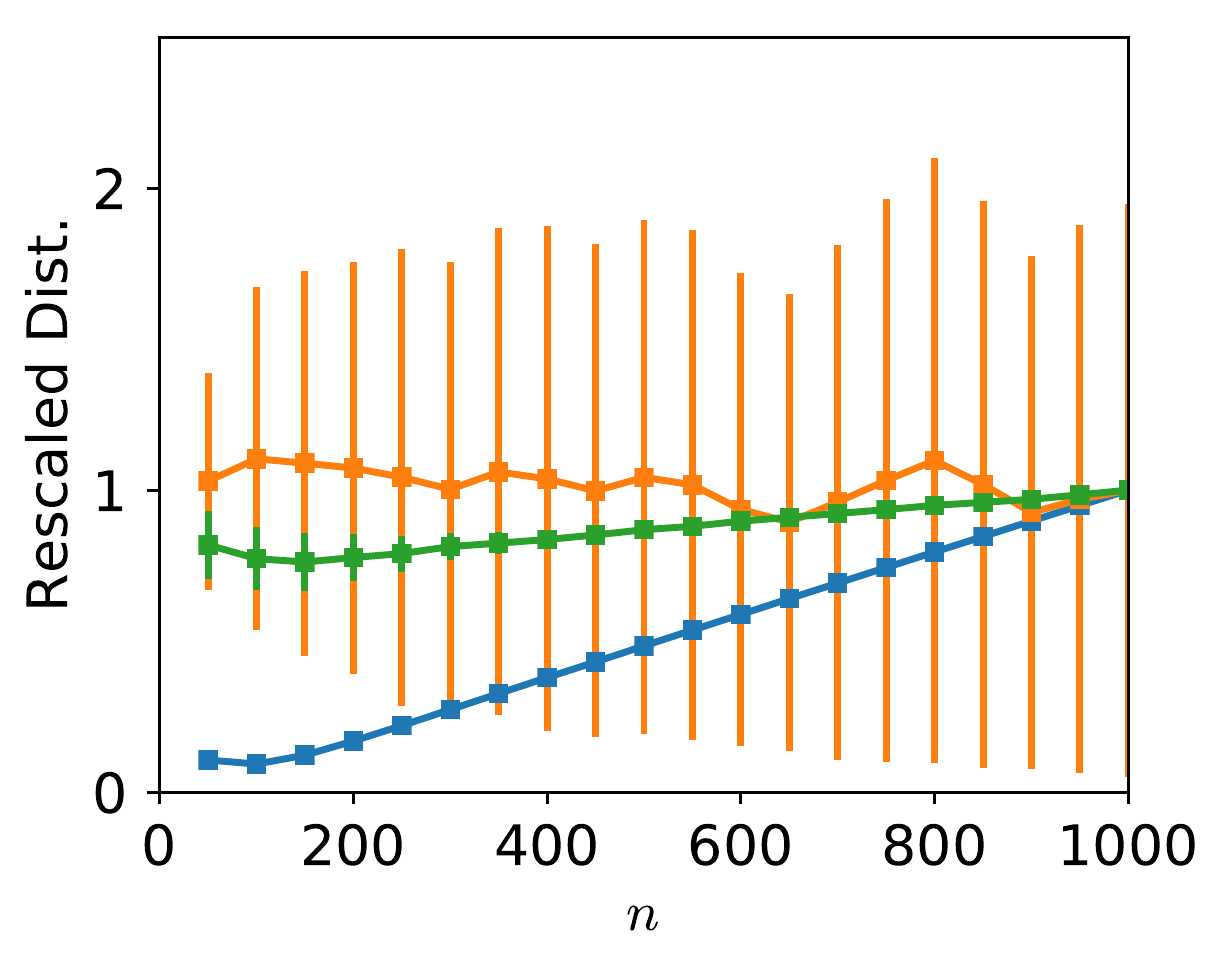}
        \caption{}
    \end{subfigure}
    \caption{
        Graph distance sensitivity for graphs generated from the same model with varying number of vertices.
        Here the models used are ER with edge connection probability $p=0.25$, WS with $k=4$ neighbour connections and rewiring probability $p=0.1$, and RR graphs with degree $d=5$.
        Distances are rescaled by the average distance $\langle d(G^*_{100}, G_{1000}^{(i)})\rangle_i$ so that models can be overlaid and trends are preserved.
        (Left) The average rescaled d-NBD $\langle d(G^*_{100}, G_n^{(i)})\rangle_i$ between a graph with $100$ vertices and graphs with identical model parameters.
        The distance is both stable and non-increasing for all graph types considered.
        (Right) Graph distance using only the largest $k$ eigenvalues. 
        Since the smallest graph is of size $50$, we must choose $k<100$ to be able to compare all graphs.
        Here the graph distance increases as $n$ increases, except for the random regular graph.
    }
    \label{fig:size_sensitivity}
\end{figure}

In order to investigate the properties of the d-NBD we consider three random graph models; the Erd\H{o}s-R\'enyi graph (ER), the Watts-Strogatz graph (WS), and the $d$-regular random graph (RR).
One of the main advantages of the d-NBD over other graph distances is that ability to compare graphs of varying size.
Figure~\ref{fig:size_sensitivity} illustrates the stability of the d-NBD for graphs of size $50$ to $1000$.
Considering a benchmark graph with $100$ vertices we subsequently calculate the distances to an ensemble of graphs all generated from the same model.
That is, we generate a reference graph $G^*_{100} ~ \rm{Model}$ and for each $n \in [50,100,\dots,1000]$ we sample $200$ graphs $(G_n^{(i)})_{i=1}^{200}$ from that model and of that size before computing $d(G^*_100, G_n^{(i)})$ for each graph.
Due to randomness in each model the distances are small but non-zero (Figure~\ref{fig:size_sensitivity}(a)) however this distance remains constant as $n$ grows.
The distances do however become noisier for small $n$ as the graph properties for smaller graphs are more susceptible to stochastic fluctuations.

In contrast, taking the largest $k$ eigenvalues of $B'$ does not provide a stable distance as $n$ varies (Figure~\ref{fig:size_sensitivity}(b)).
For both ER and WS graphs the distance from the reference graph is monotonically increasing with $n$.
The distance for the regular graph is however stable.
This can be attributed to the fact that the largest degree, which correlates with the largest eigenvalue, is fixed.
The d-NBD therefore is able to compare graphs generated from the same model, even if the degree distribution changes.

A further test we can consider on synthetic graphs is whether the d-NBD has a continuous dependence on the model parameters.
For this purpose we consider the Watt-Strogatz model since we can consider dependency is both the nearest neighbour connection parameter $k$ and the edge rewiring parameter $p$.
We can interpolate between regularised ring structure ($p=0, k \ge 2$) to completely randomised connections.
Similarly we can interpolate between sparsely connected graphs ($k=2$) to the complete graph ($k=n-1$).

\begin{figure}
    \centering
    \includegraphics[width=0.6\linewidth]{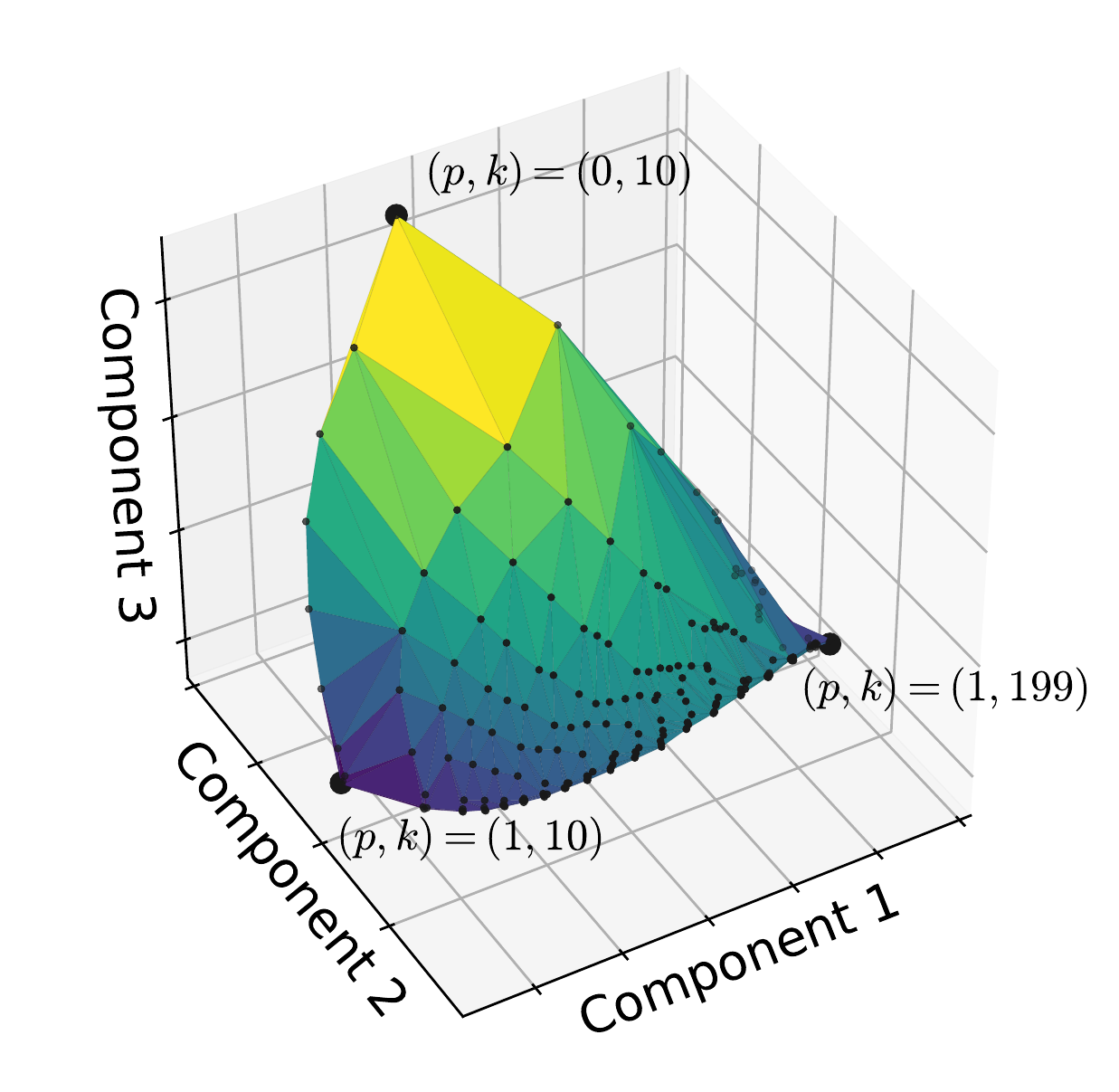}
    \caption{
        An embedding of Watts-Strogatz graphs with $200$ vertices and with rewiring parameter $p \in [0,1]$ and vertex neighbours $k \in [2, 199]$.
        Each point is the average embedding of $100$ samples from a particular model.
        The three dimensions displayed at the three principle components extracted from a $10,000$-dimensional embedding.
        Note that when $k=199$ we have a complete graph and so $(p,k)=(0,199)$ is identical to $(p,k)=(1,199)$. 
    }
    \label{fig:watts_strogatz_mesh}
\end{figure}

By considering a $10,000$-dimensional embedding of these graphs, via \eqref{eqn:embedding}, we can see that the space of Watts-Strogatz networks lies on a manifold (Figure~\ref{fig:watts_strogatz_mesh}.)
The embedding has been reduced to three dimensions using principle component analysis and each point is taken as the average position of $100$ graphs with $200$ vertices.
This observation gives further credit to the distance as changes in the underlying graph model give predictable changes in the underlying embedding.
It must be stressed that the changes in the embedding are non-trivial functions of the parameter changes.
This is clear by the uneven spacing of points in Figure~\ref{fig:watts_strogatz_mesh} which were generated by an even spacing of parameter space.

These results on synthetic graphs have shown that the d-NBD captures, but does not necessarily inform about, the generating process for these graphs and is able to detect subtle changes in the underlying model parameters. 
By their very nature, these synthetic graphs are generated from simple mechanisms.
Is the d-NBD is able to capture more complex generative mechanisms?

\subsection{Empirical Graphs}

Graph distances are useful in practice to be able to determine either the function of a graph or the (closely related) mechanisms for graph formation.
To illustrate how the d-NBD can distinguish between these mechanisms we consider a number of graph collections (detailed in Table~\ref{tab:datasets}) drawn from different fields.
\begin{table}
    \centering
    \caption{
        Empirical datasets consider in this section, in addition to three synthetic graph models for comparison.
        Each dataset contains a collection of graphs of varying size, however within each collection all graphs are assumed to be generated from the same mechanism.
        For synthetic graphs we consider graphs with mean size $200$ and variance $40$.
        ER graphs have rewiring parameter $p=0.1$., the WS graphs have $k=20$ and $p=0.1$, and the BA graphs have incoming vertices connect to $m=2$ vertices with the system initialised with a single vertex.
        All graphs are taken to be simple and undirected.
        }
    \begin{tabular}{lrrrr}
	\toprule[1pt]
	\textbf{Dataset}        & {} & \textbf{Count} & $\bm{\max(n)}$  & $\bm{\min(n)}$                 \\
    \midrule[0.5pt]
    Facebook100 (FB)       & \cite{traud2012social}    & 26    & 4563 & 769      \\
	World Subways (SW)        & \cite{roth2012long}       & 15    & 433  & 82   \\
	Autonomous Systems (AS)     & \cite{leskovec2005graphs} & 30    & 3144 & 2948\\
	Metabolic Networks (MN)     & \cite{huss2007currency}   & 10    & 1593 & 505\\
	Erd\H{o}s-R\'enyi (ER)     & \cite{newman2010networks} & 30    & 244  & 151       \\
	Watts-Strogatz (WS) & \cite{newman2010networks} & 30    & 244  & 151        \\
	Barab\'asi-Albert (BA) & \cite{newman2010networks} & 30    & 244  & 151        \\
	\bottomrule[1pt]
\end{tabular}

    \label{tab:datasets}
\end{table}
For example, we consider a sample of $26$ graphs from the Facebook100 dataset which describes the social connections within universities in the US.
These networks are human-made (although technologically assisted) and capture a number of human behaviours such as assortativity and clustering \cite{traud2012social}.
In contrast the metabolic networks dataset captures the reactions between metabolites within a number of different organisms.
These graphs are generated based on chemical and biological interaction, although arguably there has been human interference in the design and curation of such graphs.
A useful distance measure should therefore be able to distinguish between these different graph types by having small intra-type distances and relatively large inter-type distances.

\begin{figure}
    \centering
    \includegraphics[width=0.6\linewidth]{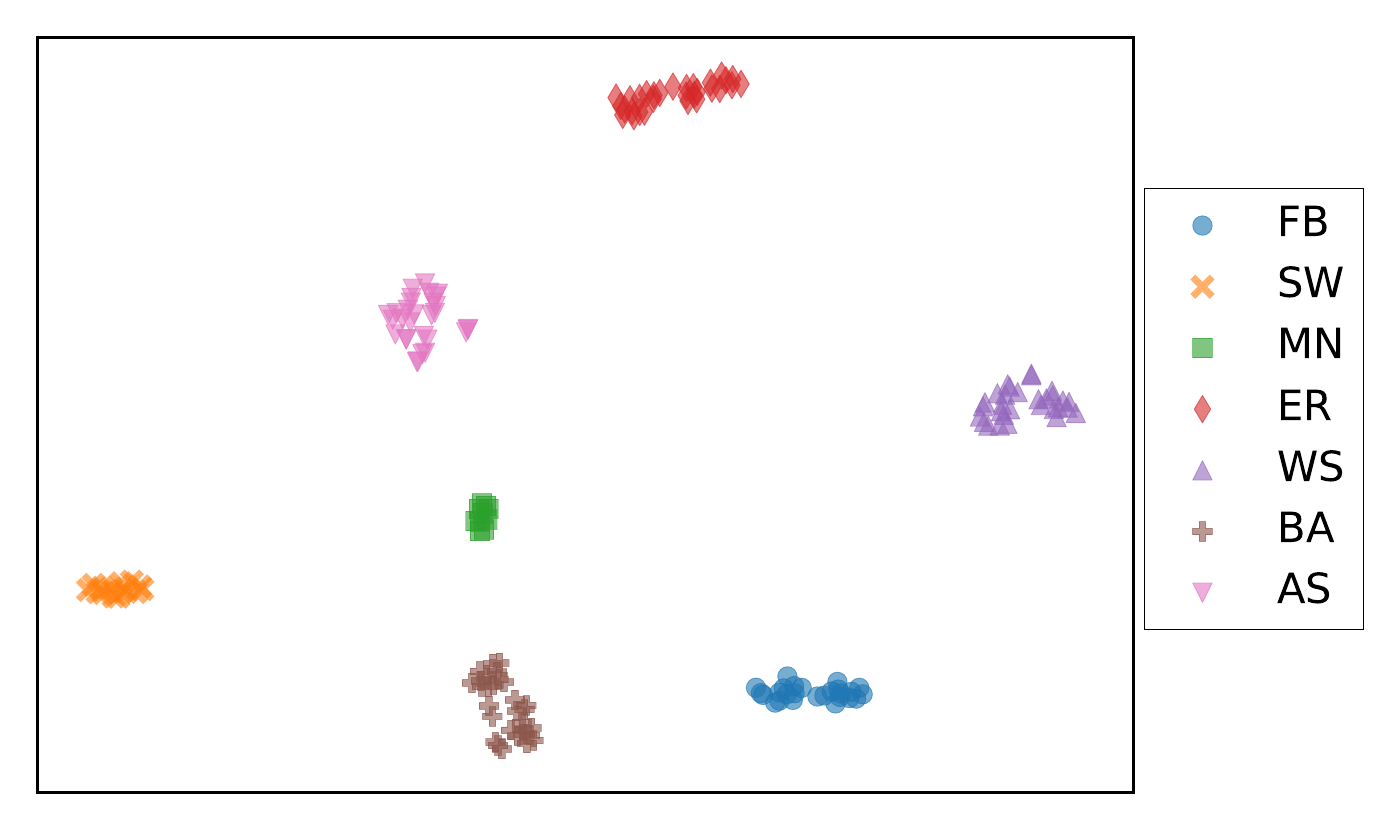}
    \caption{
        A two-dimensional projection of the spectral embedding \eqref{eqn:embedding} of the graphs listed in Table~\ref{tab:datasets}.
        The projection was created using the t-SNE algorithm \cite{maaten2008visualizing}.
        The samples from each graph collection are well clustered with zero overlap between them.
    }
    \label{fig:empirical_graphs}
\end{figure}

Figure~\ref{fig:empirical_graphs} shows a two-dimensional projection of the spectral embedding \eqref{eqn:embedding} for each graph, each of which is coloured by the graph collection it came from.
Visually it is evident that the embedding performs well as each data collection forms its own isolated cluster.
This is even the case for network collections, such as the Facebook100 graphs where the largest graph is approximately six times as large as the smallest graph in the collection.

To assess the equality of the clustering more formally we conduct a $k$-nearest neighbour classification \cite{altman1992introduction} with the number of nearest neighbours $k=10$.
We compare the clustering of the d-NBD against two benchmarks.
First, the non-backtracking spectral distance (NBD) \cite{torres2018graph} which is simply the Euclidean distance between the largest eigenvalues.
Since the smallest graph we consider has $82$ vertices we can only consider up to $164$ eigenvalues.
We also compare against the spectrum of the graph Laplacian, again taking up to $82$ of the largest eigenvalues.
Similarly to the non-backtracking matrix, the graph Laplacian (in particular the spectrum) has been previously used to characterise a graph \cite{ren2011graph}.

We assess the quality of the clustering using both the training and testing accuracy, precision, and recall.
The results are presented in Table~\ref{tab:clustering_performance}.
\begin{table}
    \centering
    \caption{
        The clustering performance of the distributional non-backtracking spectral distance (d-NBD), the non-backtracking spectral distance (NBD), and the Laplacian (LAP) distance.
        Results are averaged over a 10-fold cross-validation.
        For the graph collections considered, the d-NBD performs without fault achieving perfect accuracy, precision, and recall.
        Both the NBD and LAP clusterings  also show relatively high accuracy but a number of graphs are consistently labelled incorrectly. 
        }
\begin{tabular}{lrrr|rrr}
	\toprule[1pt]
	{}     & \multicolumn{3}{c}{\textbf{Train}} & \multicolumn{3}{c}{\textbf{Test}}                             \\
	{}     & \textbf{Rec.}                       & \textbf{Prec.}                     & \textbf{Acc.}  & \textbf{Rec.}  & \textbf{Prec.} & \textbf{Acc.}  \\
	\midrule[0.5pt]
	d-NBD  & 1.00                      & 1.00                     & 1.00 & 1.00 & 1.00 & 1.00 \\
	NBD    & 0.96                      & 0.92                     & 0.94 & 0.97 & 0.95 & 0.95 \\
	LAP     & 0.92                      & 0.80                     & 0.84 & 0.92 & 0.79 & 0.84 \\
	\bottomrule[1pt]
\end{tabular}
    \label{tab:clustering_performance}
\end{table}
To prevent overfitting we average the results over a $10$-fold cross validation.
The d-NBD achieves perfect performance for both the training and test data.
Arguably this may not be a difficult classification task and so these results should be be interpreted as a flawless measure.
However even in this `simple' task both the NBD and Laplacian methods struggle to classify a number of graphs correctly.
This is likely due to the methods not being able to utilise the full spectrum when comparing graphs of dissimilar size.

We can alternatively use more sophisticated clustering algorithms such as a random forest classifier.
These algorithms provide a slight improvement in accuracy for both NBD and the Laplacian method however neither achieve the accuracy of the d-NBD clustering.

\section{Conclusions}
\label{sec:conclusions}

In this article we have presented a method of graph comparison which can be used to compare graphs of different sizes.
Through application to both synthetic and real graphs we have shown that the d-NBD can successfully identify the differences between graphs generated by different mechanisms.
This distance performs better than distances defined on the truncated spectra of both the Laplacian and non-backtracking matrix, however this comes at a cost of having to compute the entire graph spectrum.
This is not an issue for graphs of moderate size, but can become computationally taxing for graphs formed of millions of vertices.

One possible remedy to this may be to consider the pseudospectra of such matrices.
The pseudospectra gives some information on the location of eigenvalues and can be calculated to a prescribed precision within the complex plane \cite{trefethen2005spectra}.
The pseudospectrum also has the added property that it gives a measure of the stability of the eigenvalues to small perturbations in the adjacency matrix. 
It should therefore give an indication to the effect of adding or removing an edge from the graph.

Another unexplored avenue is the use of graph distances to infer the graph generating mechanism rather than using the distance to distinguish between potentially different mechanisms.
Approaches in this area are predominantly statistical, such as the work of \citet{chen2018flexible} who use approximate Bayesian computation to select a likely graph generation mechanism.
These approaches are heavily computational especially if the number of model choices is large.
The d-NBD could be used as a precursor to such analysis by filtering possible models based on the graph distance to representative samples from these models.

The d-NBD has yet to be tested on a larger set of data which includes a wider range of graph categories (social, biological, etc.) but also with a diversity of graph collections within each category (Twitter, Facebook, and Instagram social graphs for instance).
We anticipate that the d-NBD will continue to outperform other truncated spectral methods on the basis that more information is captured in the complete spectrum.

There also lies the more fundamental question of why spectral methods work in clustering graphs (both for the Laplacian and non-backtracking operators).
The non-backtracking operator is closely related to the deformed Laplacian \cite{grindrod2018deformed}, also know as the Bethe Hessian \cite{saade2014bethe}.
This suggests there may be a physical interpretation to elucidate the underpinnings of such operators.
What does a family of Laplacian-like operators tell us about the graph structure, and how many such operators are sufficient to characterise a graph?
Connecting these two areas could lead to possible generalisations which could provide significant improvements in graph comparison.

\section*{Acknowledgments}

The authors would like to thank Renaud Lambiotte for useful discussion and both Renaud and Ebrahim Patel for a careful reading of early versions of this article.

\end{document}